\documentclass[letterpaper, 10 pt, conference]{ieeeconf}  

\IEEEoverridecommandlockouts                              
\usepackage{amsmath} 
\usepackage{amssymb}  
\usepackage{mathtools}
\usepackage{url}
\usepackage{bbm}
\usepackage{courier}
\usepackage{algorithmicx}
\usepackage{subcaption}  
\usepackage{cleveref}
\usepackage{algorithm}
\usepackage[noend]{algpseudocode}

\usepackage{xcolor}
\usepackage{siunitx}
\usepackage{booktabs}

\DeclareMathOperator*{\argmin}{arg\; min}
\DeclareMathOperator*{\argmax}{arg\; max}

\newcommand{\revmykel}[1]{{\color{black} #1}}

\title{\LARGE \bf
An Iterative Bayesian Approach for System Identification based on Linear Gaussian Models
}

\author{Alexandros E. Tzikas$^{1}$ and Mykel J. Kochenderfer$^{1}$
\thanks{$^{1}$A. E. Tzikas (corresponding author) and M. J. Kochenderfer are with the Department of Aeronautics and Astronautics, Stanford University, Stanford, CA 94305, U.S.A.
        {\tt\small \{alextzik, mykel\}@stanford.edu}}
}

\begin{document}

\maketitle
\thispagestyle{empty}
\pagestyle{empty}

\begin{abstract}

We tackle the problem of system identification, where we select inputs, observe the corresponding outputs from the true system, and optimize the parameters of our model to best fit the data. We propose a practical and computationally tractable methodology that is compatible with any system and parametric family of models. Our approach only requires input-output data from the system and first-order information of the model with respect to the parameters. 
Our approach consists of two modules. First, we formulate the problem of system identification from a Bayesian perspective and use a linear Gaussian model approximation to iteratively optimize the model's parameters. In each iteration, we propose to use the input-output data to tune the covariance of the linear Gaussian model. This online covariance calibration stabilizes fitting and signals model inaccuracy. Secondly, we define a Gaussian-based uncertainty measure for the model parameters, which we can then minimize with respect to the next selected input. We test our method with linear and nonlinear dynamics.

\end{abstract}

\section{Introduction}


In this work, we are interested in the problem of fitting a model to a system and choosing informative inputs to improve the model. Our formulation tackles the general problem of model-fitting and dataset collection, which is known as system identification. Obtaining a good model of the system is a key requirement in various engineering tasks. System identification is usually the first step in a larger pipeline that involves the control of the underlying system, e.g., with model predictive control \cite{kaiser2018sparse}. Fitting model parameters to data can be challenging when the model is a complicated function of the parameters. Ott et al. have shown that selecting control inputs based on the current model can improve the system identification process \cite{ott2024informative}.
Informative input design can reduce the data gathering requirements for system modeling, which can be costly in real-world applications. This can be particularly useful in high-dimensional problems, where finding informative directions to perturb the system can be challenging. 

There are four basic elements in the problem of system identification. We need to select a parametric family of models for the underlying system. The family must be expressive enough to capture the true system, but also simple enough to optimize over. Our parameter estimate is improved once more data becomes available. We must also choose the optimization approach that updates the model parameters. A vast literature exists on methods for estimating the model parameters \cite{pronzato2008optimal}. Furthermore, we need an uncertainty measure to minimize over the next input. This allows us to select inputs that best inform our choice of model parameters. Various objectives based on the Fisher information matrix can be minimized over the input \cite{de2017d}. This optimization problem must also be accompanied by an optimization approach.

The specifications discussed above make designing methods for system identification challenging and lead to algorithms that are usually limited to a specific set of parametric models. Our work deals with the following question:
\begin{itemize}
    \item \textit{Can we design a method for system identification that is computationally light, compatible with any parametric model family and able to determine the suitability of the model family for a given system?}
\end{itemize}
Such a method would be useful for practitioners who usually want a fast and model-agnostic approach.


Our approach leverages first-order information to construct a linear Gaussian model for the output given the input, as a function of the model parameters. We use the first-order Taylor expansion with respect to the model parameters and solve successive convex problems to improve the model parameter estimate. This simplifies the model update step discussed in the previous paragraph, which becomes computationally tractable. We also propose a procedure to tune the covariance associated with the linear Gaussian model in an online manner. This allows us to statistically calibrate our model and judge the suitability of the parametric model family. Finally, using the linear Gaussian model we define an uncertainty measure to optimize over the next input. 

The rest of the paper is organized as follows. In \cref{sec:priorwork} we discuss prior work. In \cref{sec:problem}, we formulate the problem of system identification, while in \cref{sec:proposed} we describe our proposed approach. We test our algorithm in various scenarios in \cref{sec:results} and the paper concludes with \cref{sec:conclusion}.
\section{Prior Work}\label{sec:priorwork}

The problem of system identification \revmykel{has been well studied}. \revmykel{Research} on the optimal selection of inputs 
\revmykel{dates} back to the 1960s \cite{wallis1968optimal, nahi1971design}. Recent advances in computational capabilities and learning-based methods have once again brought attention to the problem \cite{noel2017nonlinear, lee2024active}. \revmykel{This section summarizes the prior work on parameter estimation and informative input design for system identification. }

Estimating the model parameters using input-output data is usually formulated as a regression problem \cite{jennrich1969asymptotic}. A regression problem can be equivalently viewed as maximum likelihood estimation (MLE) under an appropriate distribution. For example, least squares regression corresponds to MLE with a Gaussian distribution. When estimating the model parameters, it is common to assume the true system belongs to the parametric model family \cite{pronzato2008optimal, de2017d, hjalmarsson2009system}. Then, under regularity conditions on the distribution of the input or the structure of the system, the parameter estimate is concentrated around the true system parameter \cite{pronzato2008optimal}. \textit{Our work deals with the problem of system identification when the true system does not necessarily belong to the selected parametric model family. In this case, it is critical to obtain a good fit but also a sense of the suitability of the selected model family. }

Under the assumptions mentioned in the previous paragraph, we usually obtain an unbiased estimate for the model parameters. This implies that we can design informative inputs by controlling the covariance of the estimator. Assuming an efficient estimator, the covariance converges to the inverse Fisher information matrix \cite{de2017d}. Therefore, a scalar measure of the Fisher information matrix is usually optimized over the inputs. Different scalar functions of the Fisher information matrix are used. For example, a D-optimal input maximizes the determinant of the Fisher information matrix \cite{galil1980d}. This corresponds to minimizing the uncertainty volume of the parameter estimate. \textit{Our approach optimizes any desired scalar function of an approximate Fisher information matrix to determine the next input.} Usually, the informative input design is done before the parameter estimation, but active techniques also exist \cite{mania2022active}. \textit{Our approach is a simple and generally applicable active system identification approach.}

In various applications, the system contains modes that are not properly described by a linear model. In these cases, we must deal with nonlinear models and nonlinear system identification. Early results focused on specific system dynamics for which the input design problem can either be relaxed using linear matrix inequalities \cite{hjalmarsson2007optimal} or techniques from linear systems can be applied to \revmykel{design informative} step inputs \cite{barker2004optimal}. In certain instances, the nonlinearity can be separated from the linear part of the model in what is known as structured nonlinear system identification \cite{de2017d}. There is consensus among researchers that nonlinear system identification is more difficult than linear system identification \cite{schoukens2019nonlinear}. This is, to some extent, because nonlinear optimization is more challenging than linear programming. \textit{Our approach deals with the identification of nonlinear dynamics using convex programming. Convex programs are solved efficiently using existing packages \cite{diamond2016cvxpy}.}

We examine methods for nonlinear system identification that are relevant to our approach. 
Goodwin presented an algorithm for informative input design in the general nonlinear case that is based on linearization of the parametric model \cite{goodwin1971optimal}. \textit{Our work adds to \revmykel{this work} in an important way: our method updates the estimate of the model parameters in a way that respects the linearization and is based on convex optimization, which makes it computationally fast.}
A similar work was done by Babar and Baglietto \cite{babar2016mpc}, who propose an extended Kalman filter for the joint state and parameter estimation of the dynamical system. Model predictive control techniques are used to select the next input. \textit{We improve upon their work by adding a trust-region constraint to the update of the model parameters after the linearization and by proposing a modified optimization objective for the selection of the next input that does not rely on sampling. 
} 
\textit{In contrast \revmykel{with} both works \cite{goodwin1971optimal, babar2016mpc}, we tune the noise covariance at every iteration. In addition to statistically calibrating our approach, the tuned covariance serves as a lightweight diagnostic about the suitability of the selected model family. We demonstrate this empirically.}

\section{Problem Formulation}\label{sec:problem}
We assume that a system $g$ is characterized by the relation
\begin{equation}
    y = g(x),
\end{equation}
where $x \in \mathbb{R}^{d_x}, y \in \mathbb{R}^{d_y}$ and $g: \mathbb{R}^{d_x} \rightarrow \mathbb{R}^{d_y}$. We refer to $x$ as the input and $y$ as the output. If $g$ represents dynamics, the input is usually the current state and control, which we abstract \revmykel{as} $x$, while the next state is abstracted \revmykel{as} $y$. We treat system $g$ as a black-box. We can only supply an input $x$ and obtain an output $y$ from the system.

We choose a family of functions $f(\cdot; \theta)$, \revmykel{parameterized} by $\theta \in \mathbb{R}^{d_\theta}$. At every iteration, our goal is to compute:
\begin{enumerate}
    \item $\hat{\theta}$ such that $g(\cdot) \approx f(\cdot; \hat{\theta})$ using the available data,
    \item the optimal next input to the system that decreases our uncertainty over the unknown parameter $\theta$.
\end{enumerate}
In the present work, we do not deal with the selection process for the parametric family $f(\cdot; \theta)$. Nevertheless, as we will see below, our approach allows the user to test its suitability for the underlying system.

\section{Proposed Method}\label{sec:proposed}
We present our methodology for system identification. We opt for a Bayesian treatment of the problem. We suppose that the inputs and outputs are exactly observed. Our method easily extends to the case of noisy observation of the outputs. We assume that we have already observed input-output data points $\lbrace (x_i, y_i) \rbrace_{i=1}^n$ from the system:
\begin{equation}\label{eq:data}
    y_i = g(x_i).
\end{equation} 

\revmykel{We first outline our Bayesian approach for improving the current estimate of the model parameters. We then use this Bayesian formulation in order to define an optimization problem for the next input. The components presented so far can be found in the literature \cite{goodwin1971optimal}, with the exception of the trust region in the estimate improvement step and the ability to choose the underlying probability density functions. Finally, we discuss the covariance fine-tuning, which is critical in the convergence of the method and constitutes our major contribution. }
\subsection{Parameter Estimation}\label{sec:par_est}
\revmykel{The collected data satisfy \cref{eq:data}. Because the current model $f(\cdot; \hat{\theta})$ might not match the true system $g(\cdot)$ exactly, it holds that}
\begin{equation}
    y_i = f(x_i; \hat{\theta}) + \epsilon_i,
\end{equation}
where $\epsilon_i$ is the model error for the $i$-th data-point.
We treat $\epsilon_i$ as a random quantity and choose the underlying model $\epsilon_i \sim \mathcal{N}(0, \Sigma)$ for some positive definite covariance $\Sigma$. Tuning this covariance is critical in the performance of the algorithm. Assuming conditionally independent measurements, the posterior for the hidden parameter is given by 
\begin{equation}
    p(\theta \mid y_1, \dots, y_n) \propto p(\theta) \prod_{i=1}^n p(y_i \mid \theta),
\end{equation}
where we choose the prior for $\theta$, $p(\theta)$.

The maximum a-posteriori (MAP) estimate for the parameter under these assumptions is given by 
\begin{equation}\label{eq:map}
    \begin{aligned}
        \hat{\theta}_+ &= \argmax_\theta \log p(\theta \mid y_1, \dots, y_n) \\
        &= \argmin_\theta\ -\log p(\theta)+ \dfrac{1}{2}\sum_{i=1}^n \lVert y_i - f(x_i; \theta) \rVert_{\Sigma^{-1}}^2.
    \end{aligned}
\end{equation}
In general, this can be a very difficult problem to solve because it depends on the structure of the parametric family $f$ with respect to the parameter $\theta$.  We can look for a local optimum using gradient-based techniques. We opt for an approximate treatment of the optimization problem that lends itself to a convex solver \cite{boyd2009convex}.

Suppose that the current parameter estimate is $\hat{\theta}$. 
\revmykel{For model parameters close to $\hat{\theta}$, the parametric family can be accurately approximated using a first order Taylor series.} We \revmykel{write}
\begin{equation}\label{eq: taylor}
    \begin{aligned}
    f(x; \theta) &\approx f(x; \hat{\theta}) + \dfrac{\partial f}{\partial \theta}|_{(x; \hat{\theta})} \left( \theta - \hat{\theta} \right)\\
    &= b(x; \hat{\theta}) + C(x; \hat{\theta})\theta
    \end{aligned}
\end{equation}
for $\theta$ close to $\hat{\theta}$, where 
\begin{equation}
\begin{aligned}
    \left[ \dfrac{\partial f}{\partial \theta}|_{(x; \hat{\theta})} \right]_{i, j} &= \dfrac{\partial f_i}{\partial \theta_j}|_{(x; \hat{\theta})},\\
    b(x; \hat{\theta}) &= f(x; \hat{\theta}) - \dfrac{\partial f}{\partial \theta}|_{(x; \hat{\theta})}\hat{\theta},\\
    C(x; \hat{\theta}) &= \dfrac{\partial f}{\partial \theta}|_{(x; \hat{\theta})}.
\end{aligned}
\end{equation}
Plugging \eqref{eq: taylor} into \eqref{eq:map} and imposing a trust region constraint, we get that
\begin{equation}\label{eq:taylor_objective}
\begin{aligned}
    \hat{\theta}_+ \approx \argmin_{\lVert \theta - \hat{\theta}\rVert \leq \delta}\ &-\log p(\theta)
    +\\ &\dfrac{1}{2}\sum_{i=1}^n \lVert \underbrace{y_i - b(x_i;\hat{\theta})}_{\Tilde{y}_i} - C(x_i; \hat{\theta})\theta \rVert_{\Sigma^{-1}}^2.
    \end{aligned}
\end{equation}
Assuming a log-concave $p(\theta)$, this is a convex problem with respect to $\theta$ and can easily be handled by a computational framework for convex optimization \cite{diamond2016cvxpy}. The constraint $\lVert \theta - \hat{\theta}\rVert \leq \delta$ guarantees that \revmykel{we search for the next model parameter in the region where the Taylor series for $f(\cdot; \theta)$ is sufficiently close to $f(\cdot; \hat{\theta})$}. It also does not allow our next estimate to differ much \revmykel{relative to} the current one. This is justified \revmykel{because} we expect our model parameter estimate to vary smoothly across iterations. In other words, placing too much trust on any one datapoint can lead to excessively large steps or premature convergence \cite{kochenderfer2019algorithms}. 

We can replace Gaussian densities with any other log concave density and still obtain a convex problem. The used density for $\epsilon_i$ can be changed in each estimation step, based on the realized errors $\lbrace y_i - f(x_i; \hat{\theta}) \rbrace_{i=1}^n$. This allows us to select the family of underlying distributions. More details on this will be given in section \ref{sec:finetune}.

\subsection{Informative Input Design}\label{sec:inform_inp}
After obtaining $\hat{\theta}_+$, our goal is to select $x_{n+1}$. The goal of selecting $x_{n+1}$ is to reduce our uncertainty in the posterior estimate of $\theta$. Assuming that we are given a new data-point $(x_{n+1}, y_{n+1})$ and under linear Gaussian assumptions (for the prior as well), the posterior inverse covariance is given by the Hessian of the objective of \eqref{eq:taylor_objective} but evaluated using the most up-to-date MAP estimate $\hat{\theta}_+$, i.e,
\begin{equation}\label{eq:posterior_cov_taylor}
    \Sigma_\mathrm{posterior}^{-1} = \Sigma_\mathrm{prior}^{-1}+\sum_{i=1}^{n+1} C(x_i; \hat{\theta}_+)^\top \Sigma^{-1} C(x_i; \hat{\theta}_+),
\end{equation}
where we assume $\theta \sim \mathcal{N}(\theta_\mathrm{prior}, \Sigma_\mathrm{prior})$. In the case of a uniform prior for $\theta$, we can set $\Sigma_\mathrm{prior}^{-1} = 0$, i.e., we can simply ignore the prior term in \eqref{eq:posterior_cov_taylor}.

It has been shown that the covariance \eqref{eq:posterior_cov_taylor} can be accurate in the nonlinear case, although it is exact only in the linear regime \cite{schalow1967limits}. Eq.~\eqref{eq:posterior_cov_taylor} is an approximate Fisher information matrix, where a prior is added and the expectation over the data is not used. We note that it only depends on the inputs $x_1, \dots, x_{n+1}$ and not on the corresponding outputs.

To minimize the posterior covariance, we maximize some notion of magnitude, 
\begin{equation}
\mathcal{M} \left( \Sigma_\mathrm{prior}^{-1}+\sum_{i=1}^{n+1} C(x_i; \hat{\theta}_+)^\top \Sigma^{-1} C(x_i; \hat{\theta}_+) \right),
\end{equation}
of the posterior inverse covariance over $x_{n+1}$. \revmykel{We can do so because the uncertainty in the parameter estimate is proportional to the magnitude of the posterior covariance, which is inversely proportional to the magnitude of the inverse posterior covariance. For example, we can use $\log \det$ as $\mathcal{M}$. The determinant of the covariance equals the product of its eigenvalues, which are proportionally related to the semi-axes of the probability contours for a Gaussian distribution.} 

In the process of choosing $x_{n+1}$, there usually exist constraints. \revmykel{For example, }the input \revmykel{might need} to belong \revmykel{to} a specific set, $x_{n+1} \in \mathcal{X}$. The same might hold for the corresponding output, $g(x_{n+1}) \in \mathcal{Y}$. We can take such constraints into account by solving the following optimization problem:
\begin{equation}\label{eq:input_selection}
\begin{aligned}
\argmax_{x_{n+1}}\
&\mathcal{M} \left( \Sigma_\mathrm{prior}^{-1}+\sum_{i=1}^{n+1} C(x_i; \hat{\theta}_+)^\top \Sigma^{-1} C(x_i; \hat{\theta}_+) \right)\\
&+ \lambda \left(\mathbbm{1}(x_{n+1}\in \mathcal{X}) + \mathbbm{1}(f(x_{n+1}; \hat{\theta}_+) \in \mathcal{Y}) \right).
\end{aligned}
\end{equation}
In order to solve \cref{eq:input_selection} \revmykel{with} gradient descent methods, we can replace the indicators with smooth functions. By increasing parameter $\lambda \geq 0$, we can move the search towards feasible inputs $x_{n+1}$. This is an example of a penalty method \cite{kochenderfer2019algorithms}. As a receding horizon extension, we can easily extend \cref{eq:input_selection} to optimize over the next $k$ inputs, $x_{n+1}, \dots, x_{n+k}$. In \eqref{eq:input_selection}, because the system $g$ is only accessible via a query oracle, we approximate $g(x_{n+1}) \in \mathcal{Y}$ by $f(x_{n+1}; \hat{\theta}_+) \in \mathcal{Y}$.


\subsection{Fine-tuning}\label{sec:finetune}
We present the fine-tuning approach for the covariance $\Sigma$ and also propose a way to tune $\delta$. 

At the current iteration\revmykel{,} we have access to the dataset $\lbrace (x_i, y_i)\rbrace_{i=1}^n$. After obtaining the new parameter estimate $\hat{\theta}_+$, we can compute the estimated noise realizations $\epsilon^\mathrm{model}_i$ by comparing the true output to the model output, i.e., for $i=1,\dots, n$ 
\begin{equation}
    \epsilon^\mathrm{model}_{i} =: y_i - f(x_i; \hat{\theta}_+).
\end{equation}
Using these samples, we can compute a covariance for the errors caused by the mismatch between $g(\cdot)$ and $f(\cdot; \hat{\theta}_+)$:
\begin{equation}\label{eq:update_meas_cov}
    \Sigma_\mathrm{model\ error} =: \dfrac{1}{n} \sum_{i=1}^n \epsilon^\mathrm{model}_i {\epsilon^\mathrm{model}_i}^\top.
\end{equation}
However, in problem \eqref{eq:taylor_objective} we consider the distribution of the errors
\begin{equation}
    y_i - b(x_i;\hat{\theta}) - C(x_i; \hat{\theta})\hat{\theta}_+
\end{equation}
that consist of two terms: the error from the mismatch between $f(\cdot; \theta)$ and $g(\cdot)$ and the error because of the linearization, since
\begin{equation}
\begin{aligned}
    y_i - b(x_i;\hat{\theta}) - C(x_i; \hat{\theta})\hat{\theta}_+ = \\
    \epsilon^\mathrm{model}_i + \underbrace{f(x_i; \hat{\theta}_+) - \left( b(x_i;\hat{\theta}) + C(x_i; \hat{\theta})\hat{\theta}_+\right)}_{\epsilon^\mathrm{lin}_i}.
\end{aligned}
\end{equation}
This implies that the covariance $\Sigma$ should be set using the sample covariance of $\lbrace \epsilon^\mathrm{model}_i + \epsilon^\mathrm{lin}_i \rbrace_{i=1}^n$:
\begin{equation}\label{eq:SigmaUpdate}
    \Sigma \leftarrow \dfrac{1}{n} \sum_{i=1}^n \left( \epsilon^\mathrm{model}_i + \epsilon^\mathrm{lin}_i\right)\left( \epsilon^\mathrm{model}_i + \epsilon^\mathrm{lin}_i\right)^\top,
\end{equation}
where $\epsilon^\mathrm{lin}_i$ is the error between the value from the model and the value from the Taylor expansion of the model for the $i$-th output.

For setting the trust region $\delta$, we can use a simple heuristic. If the average linearization error size $n^{-1}\sum_{i=1}^n \lVert \epsilon^\mathrm{lin}_i \rVert $ becomes comparable to the average absolute model error size $n^{-1}\sum_{i=1}^n \lVert \epsilon^\mathrm{model}_i \rVert $, we reduce $\delta$. This procedure guarantees that we keep any error due to linearization small compared to the errors because of the model mismatch. This allows for confident steps in the parameter update.


\subsection{Algorithm}\label{sec:algo}
We combine the modules presented above in \Cref{alg:proposed}. We assume we have already acquired dataset $\lbrace (x_i, y_i)\rbrace_{i=1}^n$ and show the proposed procedure for augmenting the dataset by one point. 

Our algorithm first runs a loop to improve the parameter estimate. We obtain a new parameter estimate by solving problem \eqref{eq:taylor_objective}. We only accept the new solution if the linearization errors are small compared to the model errors. In this case, we update the covariance $\Sigma$ and re-solve problem \eqref{eq:taylor_objective} using the new covariance and the Taylor expansion around the new parameter estimate. If the linearization errors are deemed significant we reject the new estimate, reduce $\delta$ and re-solve. 
This iterative technique will produce a new parameter estimate.
It will further produce an error covariance $\Sigma_\mathrm{model\ error}$ for the errors attributed to the discrepancy between the current model and the true system.

In the second phase, we obtain the next input by optimizing \cref{eq:input_selection}. Note that we use $\Sigma_\mathrm{model\ error}$, because we aim to reduce the uncertainty associated with the model mismatch.

The algorithm outputs an updated $\Sigma$, which is used as input in the next algorithm call. 

\Cref{alg:proposed} allows for testing whether the chosen parametric model family $f$ is suitable for the system under consideration.  We know that the parameter estimates obtained from the successive calls to \Cref{alg:proposed} correspond to the best model fits for the obtained data-points and the linearized MAP objective. Suppose that the magnitude of $\Sigma_\mathrm{model\ error}$ does not decrease between successive calls to \Cref{alg:proposed}. This suggests that the chosen parametric family $f$ is not appropriate for the system.

\begin{algorithm}
\caption{Proposed Algorithm for Active System Identification}\label{alg:proposed}
\begin{algorithmic}[1]
\Require dataset $\lbrace (x_i, y_i)\rbrace_{i=1}^n$, parameter estimate $\hat{\theta}$, covariances $\Sigma_\mathrm{prior}$ and $\Sigma$, model family $f$, 
constraint sets $\mathcal{X}, \mathcal{Y}$, constants $\delta$ and $\lambda$

\Repeat
\State Obtain updated estimate $\hat{\theta}_+$ \revmykel{by} \cref{eq:taylor_objective}
\State Compute $\epsilon^\mathrm{model}_i, \epsilon^\mathrm{lin}_i$ for all $i=1, \dots, n$
\If  {$n^{-1}\sum_{i=1}^n \lVert \epsilon^\mathrm{lin}_i \rVert$ large compared to $ n^{-1}\sum_{i=1}^n \lVert \epsilon^\mathrm{model}_i \rVert $}
\State Reduce $\delta$
\Else
\State Update $\Sigma$ using \cref{eq:SigmaUpdate}
\State Set $\Sigma_\mathrm{model\ error} = \dfrac{1}{n} \sum_{i=1}^n \epsilon^\mathrm{model}_i {\epsilon^\mathrm{model}_i}^\top$
\State $\hat{\theta} \leftarrow \hat{\theta}_+$
\EndIf
\Until{convergence}
\State Obtain next input $x_{n+1}$ by solving \cref{eq:input_selection} where $\Sigma \leftarrow \Sigma_\mathrm{model\ error}$
\State Obtain output $y_{n+1}$ for input $x_{n+1}$ using the true system
\State \Return dataset $\lbrace (x_i, y_i)\rbrace_{i=1}^{n+1}$, $\hat{\theta}_+$, $\Sigma$, and $\Sigma_\mathrm{model\ error}$

\end{algorithmic}
\end{algorithm}
\section{Results}\label{sec:results}
We test our \revmykel{algorithm} \revmykel{on} both linear and nonlinear systems.\footnote{The code is included in the repository: \url{https://github.com/alextzik/informative_input_design}} Our results show the importance of \revmykel{fine-tuning the covariances $\Sigma$} in our algorithm.

\subsection*{Set-up}
\revmykel{We set} $\delta=0.3$. In line 5 of \cref{alg:proposed}, we reduce $\delta$ by setting $\delta \leftarrow 0.8\delta$. 
The first data-points are assumed fixed. Furthermore, we independently run the algorithm for $30$ different initializations of $\Sigma$, $\theta_\mathrm{prior}$, and $\Sigma_\mathrm{prior}$. We use the scalar function $\log \det$ as $\mathcal{M}$ when determining the next input. The loop in lines 1-10 in \Cref{alg:proposed} is carried out for $10$ times.

\subsection*{Metrics}
In \cref{sec:lin}, \cref{sec:henon}, and \cref{sec:unicycle} below\revmykel{,} the true system belongs to the parametric model family, i.e., $g(\cdot) = f(\cdot; \theta_\mathrm{true})$ for some $\theta_\mathrm{true}\in \mathbb{R}^{d_\theta}$. We can thus compare the methods with respect to the error $\lVert \hat{\theta} - \theta_\mathrm{true} \rVert_\infty$ as a function of iteration number. The iteration equals the size of the dataset.

\subsection{Linear System}\label{sec:lin}
We first consider a linear system
\begin{equation}
    g(x_t) = \begin{bmatrix}
        1 & 2\\
        3 & 4
    \end{bmatrix}x_t
\end{equation}
and the parametric model
\begin{equation}\label{eq:lin_model}
    f(x_t; \theta) = \begin{bmatrix}
        \theta_1 & \theta_2\\
        \theta_3 & \theta_4
    \end{bmatrix}x_t.
\end{equation}

\begin{figure*}[t]
    \centering
    \begin{subfigure}[b]{0.34\textwidth}
        \centering
        \includegraphics[width=\textwidth]{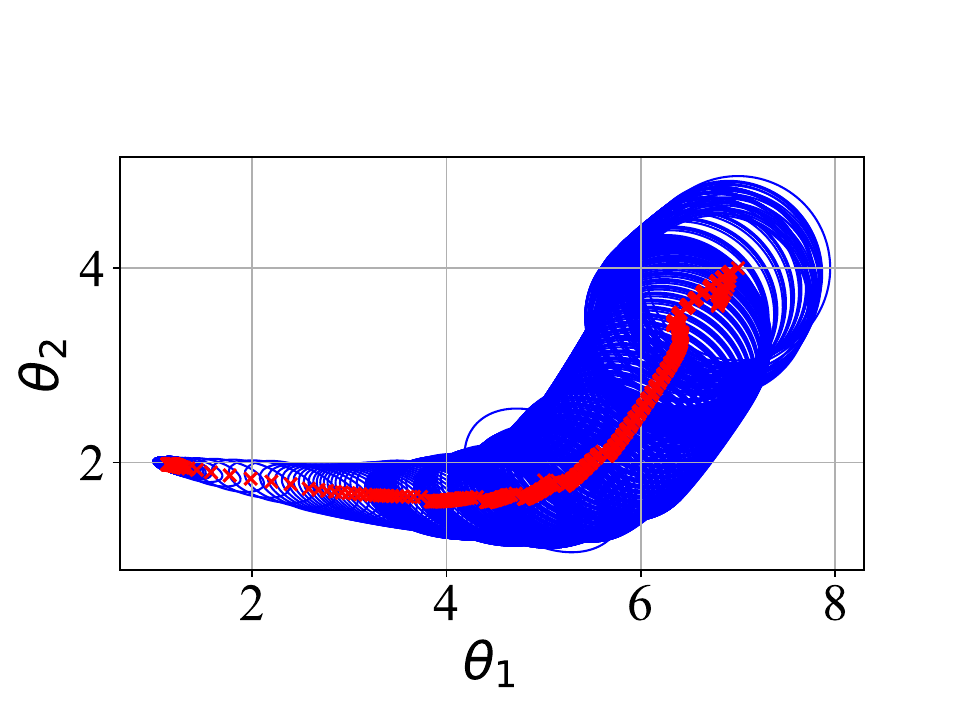}
        \caption{Model parameter estimate (point value in red and posterior covariance \eqref{eq:posterior_cov_taylor} in blue) as it evolves over time for specific $\theta_\mathrm{prior}$, $\Sigma_\mathrm{prior}$ and initial $\Sigma$.}
        \label{fig:linear_trajectory}
    \end{subfigure}
    \hfill
    \begin{subfigure}[b]{0.24\textwidth}
        \centering
        \includegraphics[width=\textwidth]{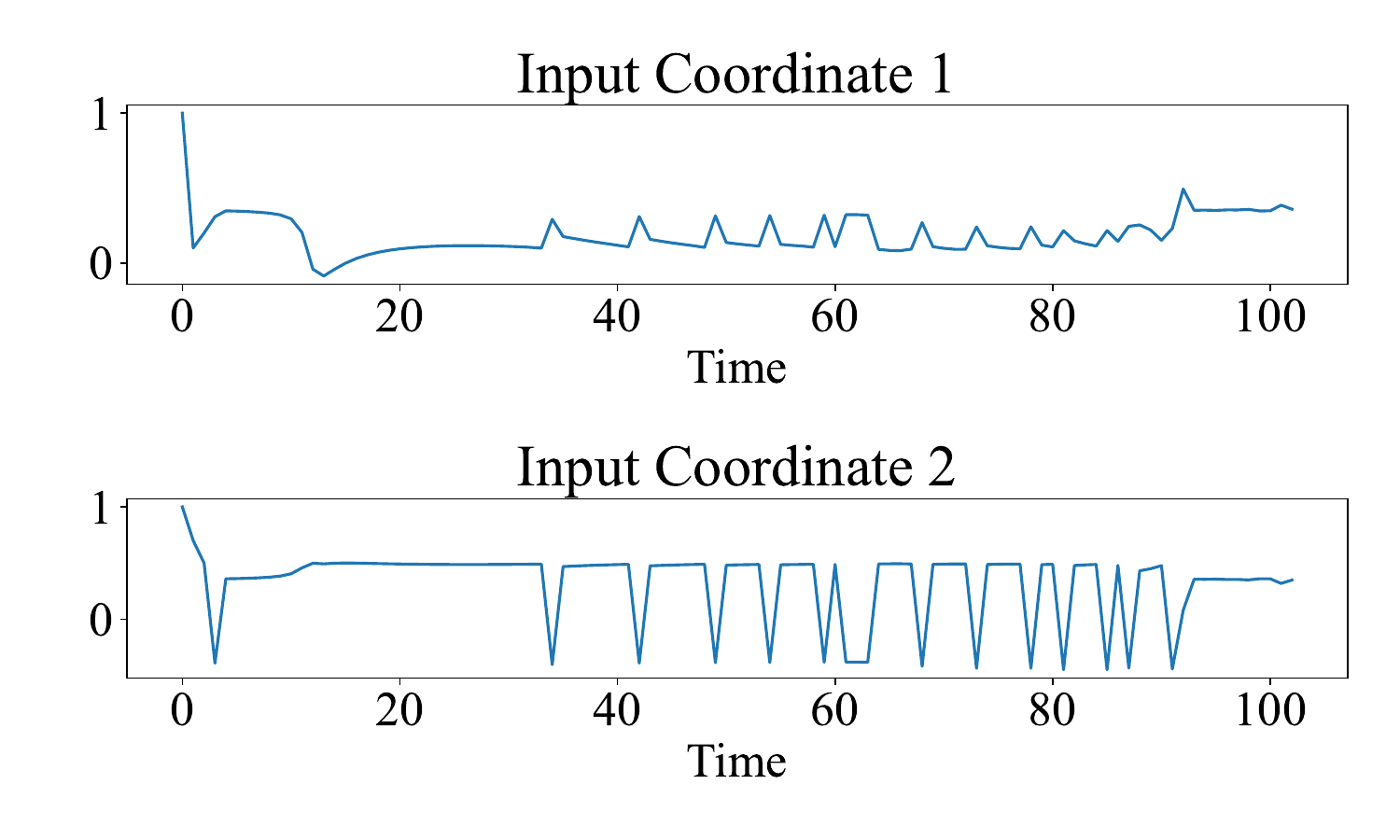}
        \caption{Inputs as a function of time for specific $\theta_\mathrm{prior}$, $\Sigma_\mathrm{prior}$ and initial $\Sigma$.}
        \label{fig:linear_input}
    \end{subfigure}
    \hfill
    \begin{subfigure}[b]{0.34\textwidth}
        \centering
        \includegraphics[width=\textwidth]{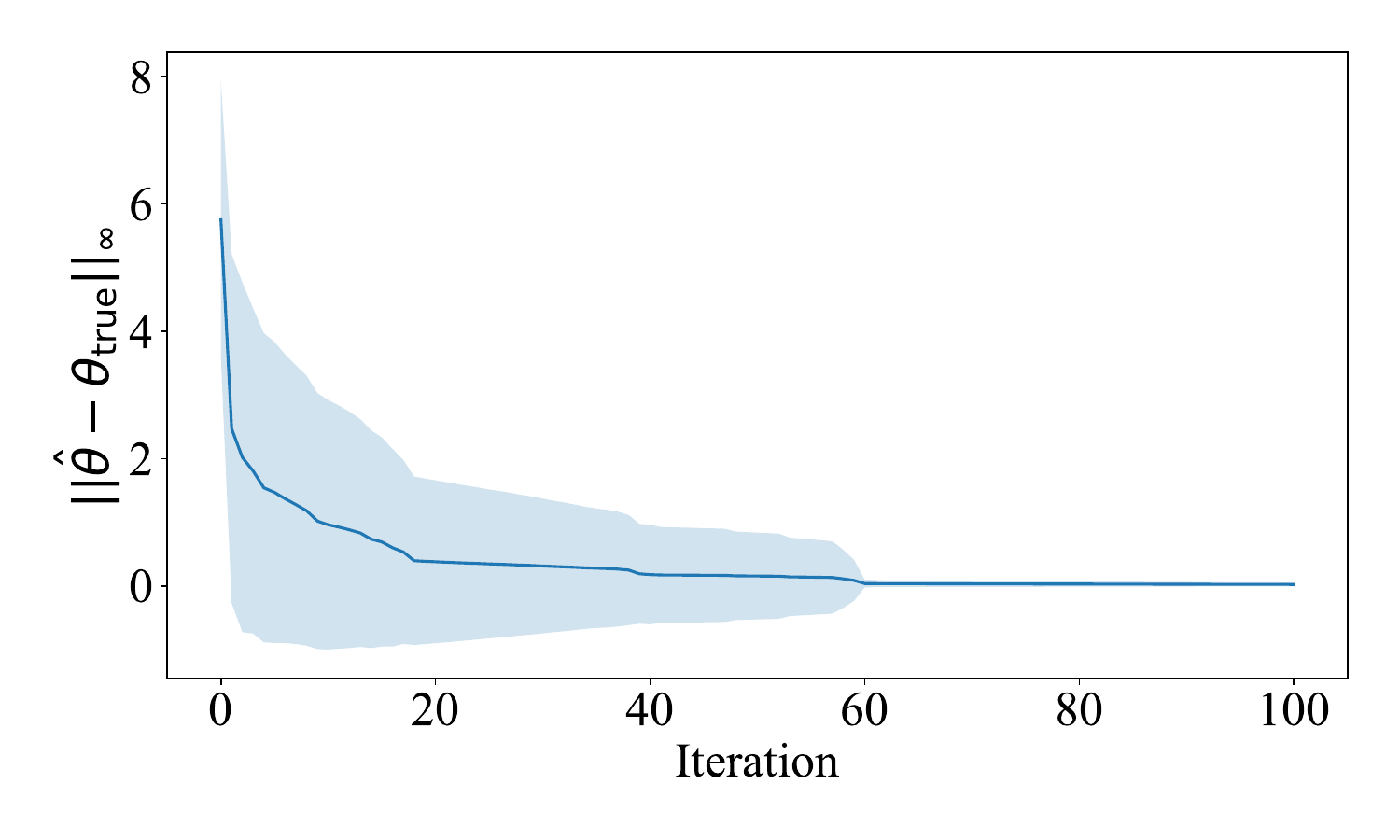}
        \caption{Mean model parameter estimate error and 1-standard deviation interval. The randomness is over $\theta_\mathrm{prior}$, $\Sigma_\mathrm{prior}$ and initial $\Sigma$.}
        \label{fig:linear_error}
    \end{subfigure}
    \caption{Results for the linear dynamics.}
    \label{fig:linear}
\end{figure*}

Our results are shown in Figure \ref{fig:linear}. In this experiment, we also bound the norm of the input to be less than $0.5$. Based on the error plot, our algorithm is able to converge to the true parameter values irrespective of the initial $\Sigma$, $\theta_\mathrm{prior}$, and $\Sigma_\mathrm{prior}$. 

In addition to the error, we plot the trajectory of the parameter estimate along with its error ellipses for a particular choice of $\theta_\mathrm{prior}, \Sigma_\mathrm{prior}$, and initial $\Sigma$. We only show the estimate\revmykel{d} trajectory of the first two coordinates of $\hat{\theta}$, $\hat{\theta}_1$ and $\hat{\theta}_2$, although the complete estimate is 4-dimensional. We further plot the designed inputs for the same $\theta_\mathrm{prior}, \Sigma_\mathrm{prior}$, and initial $\Sigma$. 

\subsection{Hénon map dynamics}\label{sec:henon}
We consider the Hénon map dynamics 
\begin{equation}\label{eq:henon}
    g(x_t) = \begin{bmatrix} 
1 - \alpha x_{t, 1}^2 + x_{t, 2} \\
\beta x_{t,1}
\end{bmatrix},
\end{equation}
where $\alpha=1.4$ and $\beta=0.3$. It is one of the most studied systems with chaotic behavior \cite{henon2004two}.
The parametric model is 
\begin{equation}\label{eq:henon_model}
    f(x_t; \theta) = \begin{bmatrix} 
1 - \theta_1 x_{t, 1}^2 + x_{t, 2} \\
\theta_2 x_{t,1}
\end{bmatrix}.
\end{equation}

\begin{figure*}[t]
    \centering
    \begin{subfigure}[b]{0.34\textwidth}
        \centering
        \includegraphics[width=\textwidth]{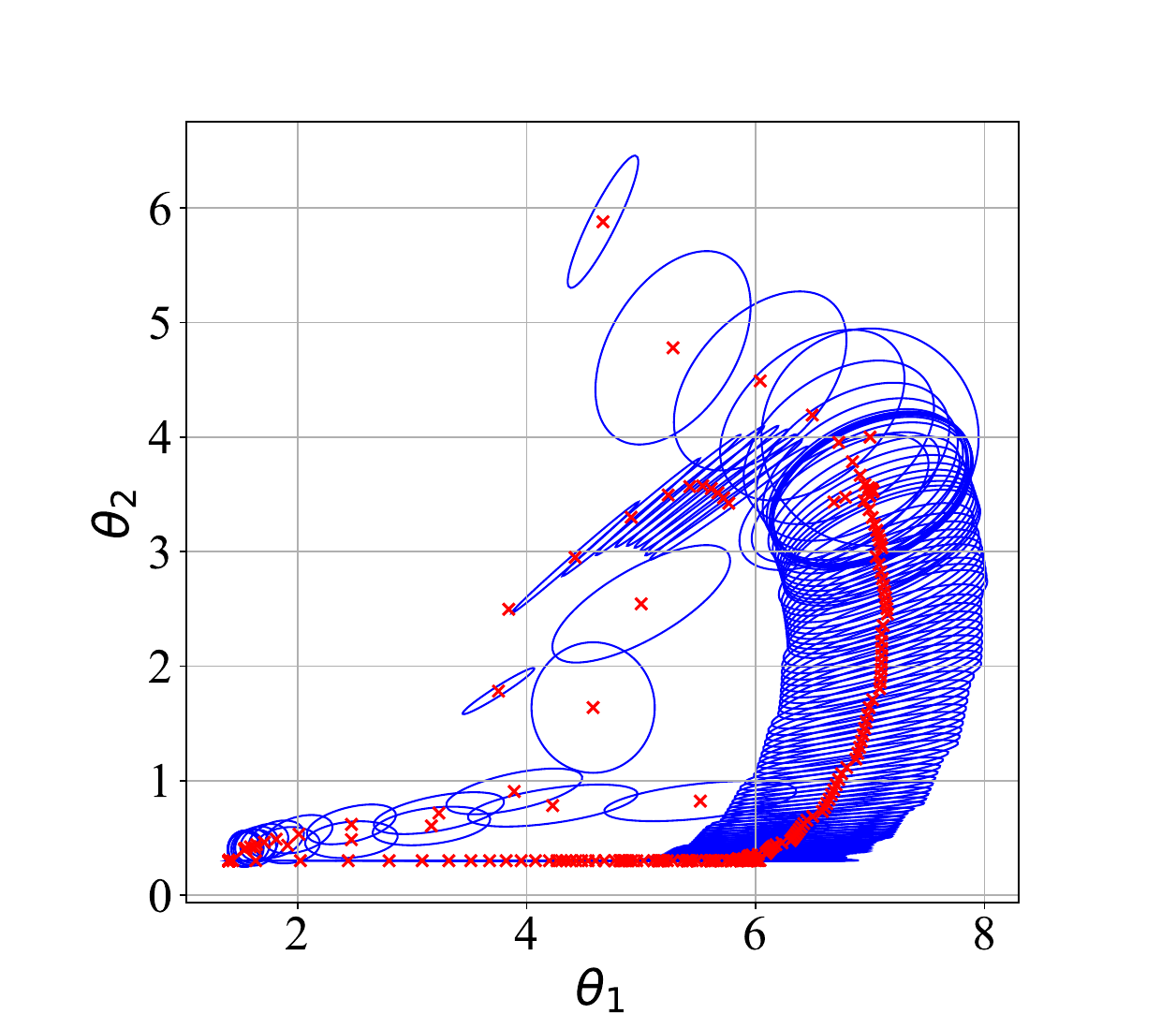}
        \caption{Model parameter estimate (point value in red and posterior covariance \eqref{eq:posterior_cov_taylor} in blue) as it evolves over time for specific $\theta_\mathrm{prior}$, $\Sigma_\mathrm{prior}$ and initial $\Sigma$.}
        \label{fig:henon_trajectory}
    \end{subfigure}
    \hfill
    \begin{subfigure}[b]{0.3\textwidth}
        \centering
        \includegraphics[width=\textwidth]{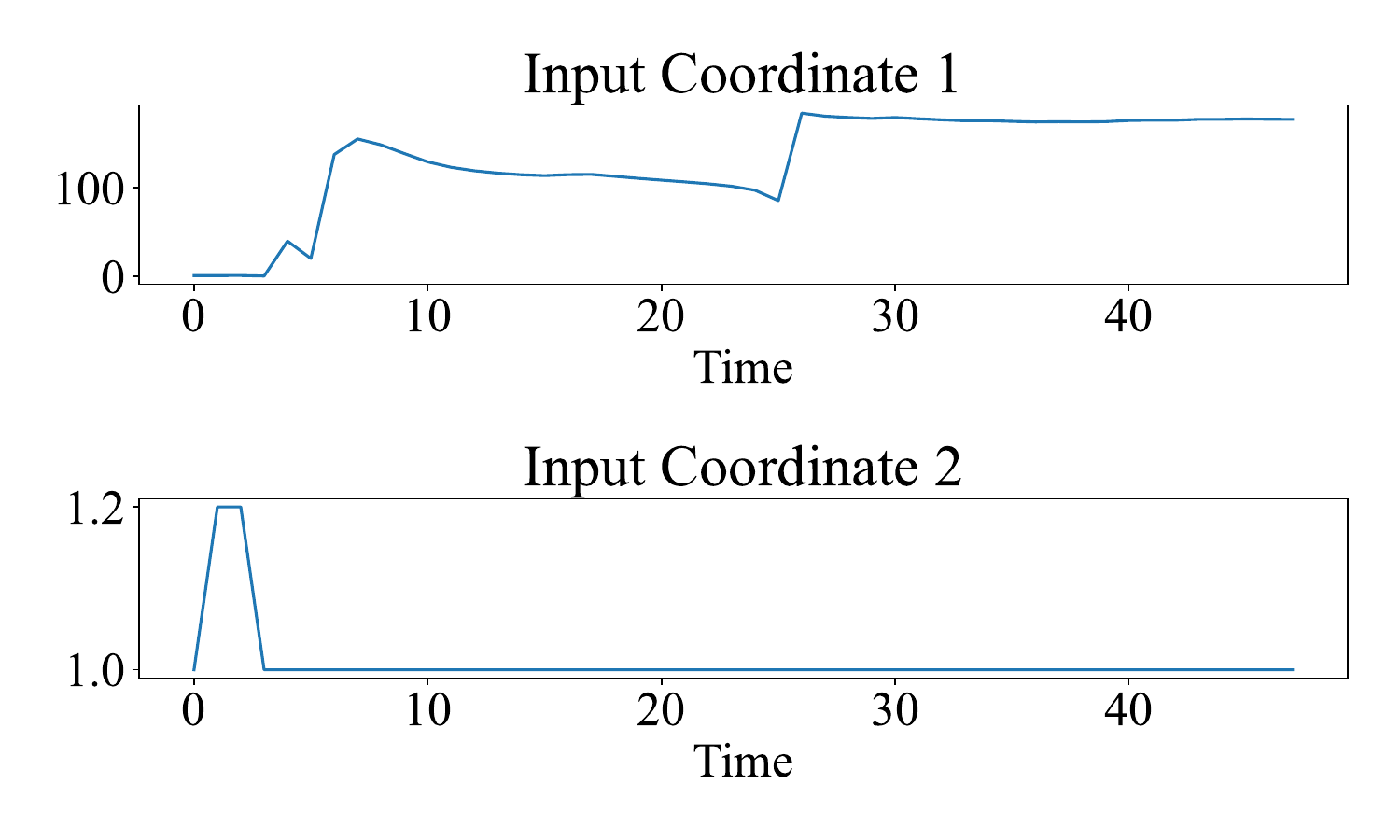}
        \caption{Inputs as a function of time for specific $\theta_\mathrm{prior}$, $\Sigma_\mathrm{prior}$ and initial $\Sigma$.}
        \label{fig:henon_input}
    \end{subfigure}
    \hfill
    \begin{subfigure}[b]{0.34\textwidth}
        \centering
        \includegraphics[width=\textwidth]{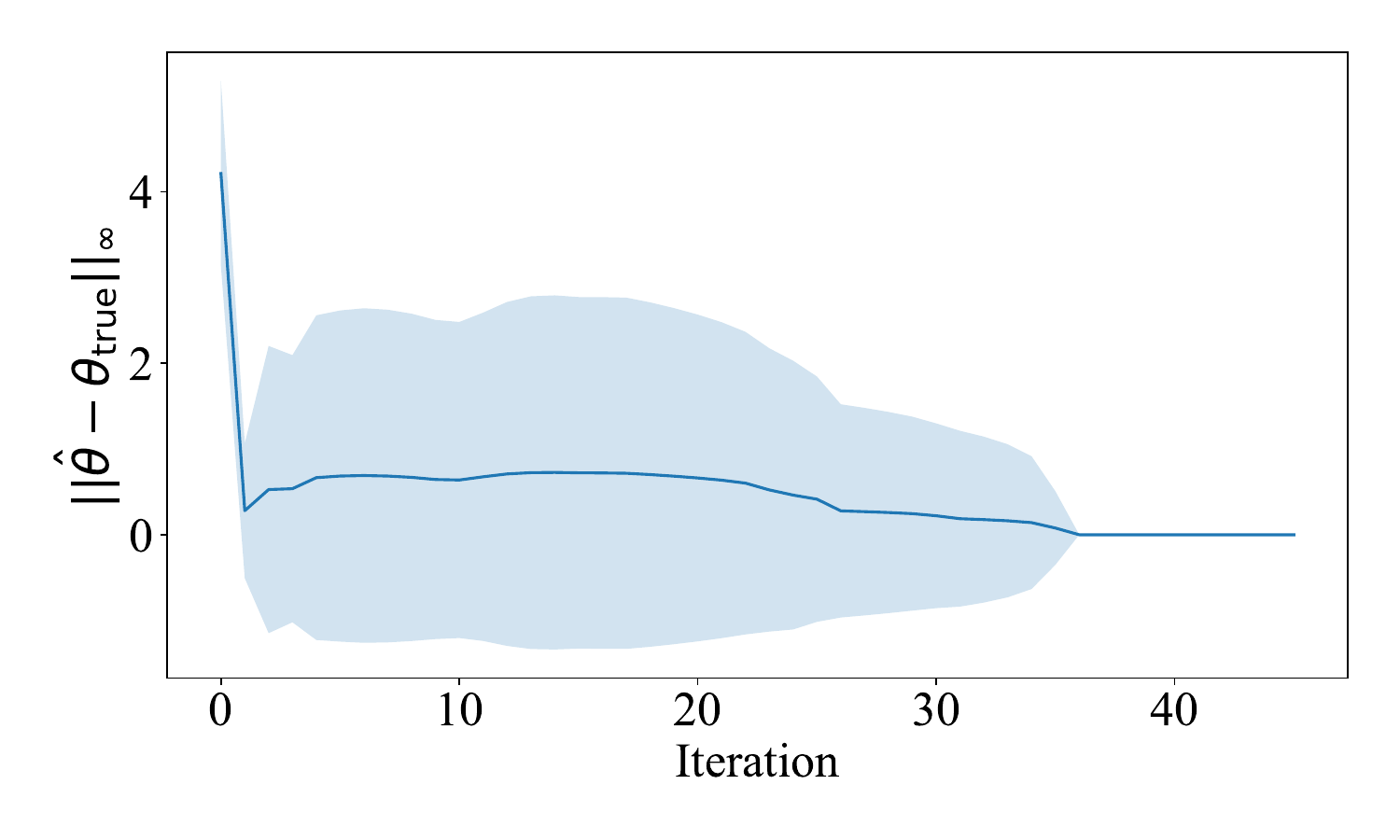}
        \caption{Mean model parameter estimate error and 1-standard deviation interval. The randomness is over $\theta_\mathrm{prior}$, $\Sigma_\mathrm{prior}$ and initial $\Sigma$.}
        \label{fig:henon_error}
    \end{subfigure}

    \caption{Results for the Hénon dynamics.}
    \label{fig:henon}
\end{figure*}

Our results are shown in Figure \ref{fig:henon}. 
Our proposed algorithm is able to converge to the true model parameters. From the estimate trajectory, we observe that the posterior covariance \eqref{eq:posterior_cov_taylor} captures the uncertainty in the parameter: based on the true parameter value, the estimate has larger errors in the first coordinate $\theta_1$ at later times, which is captured by the elongated ellipse axis along that direction for those times. We also observe that the error ellipses shrink as we move towards the true parameters. This implies that our uncertainty is reduced.

\subsection{Unicycle Model}\label{sec:unicycle}
\begin{figure*}[t]
    \centering
    \begin{subfigure}[b]{0.34\textwidth}
        \centering
        \includegraphics[width=\textwidth]{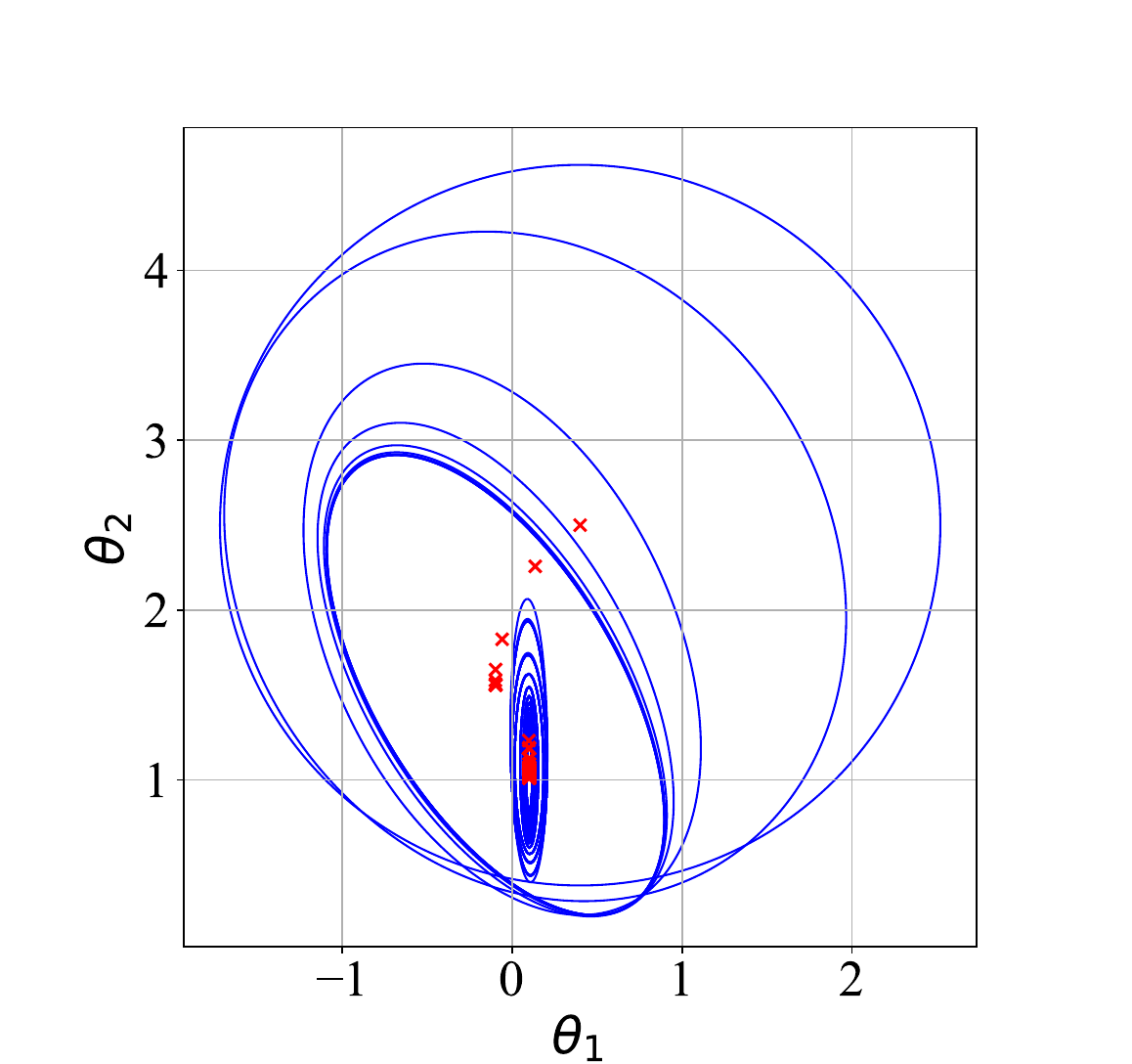}
        \caption{Model parameter estimate (point value in red and posterior covariance \eqref{eq:posterior_cov_taylor} in blue) as it evolves over time for specific $\theta_\mathrm{prior}$, $\Sigma_\mathrm{prior}$ and initial $\Sigma$.}
        \label{fig:unicycle_trajectory}
    \end{subfigure}
    \hfill
    \begin{subfigure}[b]{0.3\textwidth}
        \centering
        \includegraphics[width=\textwidth]{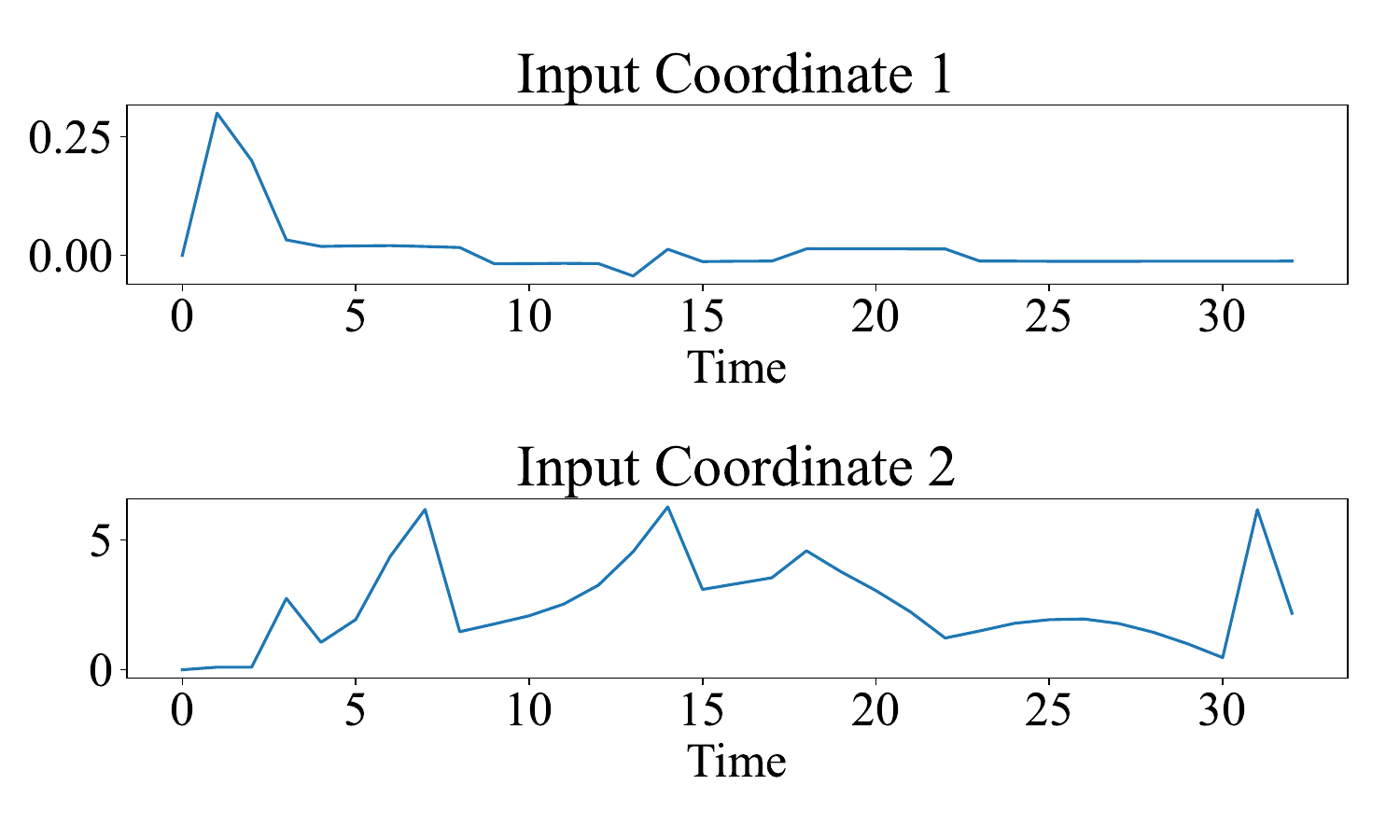}
        \caption{Inputs as a function of time for specific $\theta_\mathrm{prior}$, $\Sigma_\mathrm{prior}$ and initial $\Sigma$.}
        \label{fig:unicycle_input}
    \end{subfigure}
    \hfill
    \begin{subfigure}[b]{0.34\textwidth}
        \centering
        \includegraphics[width=\textwidth]{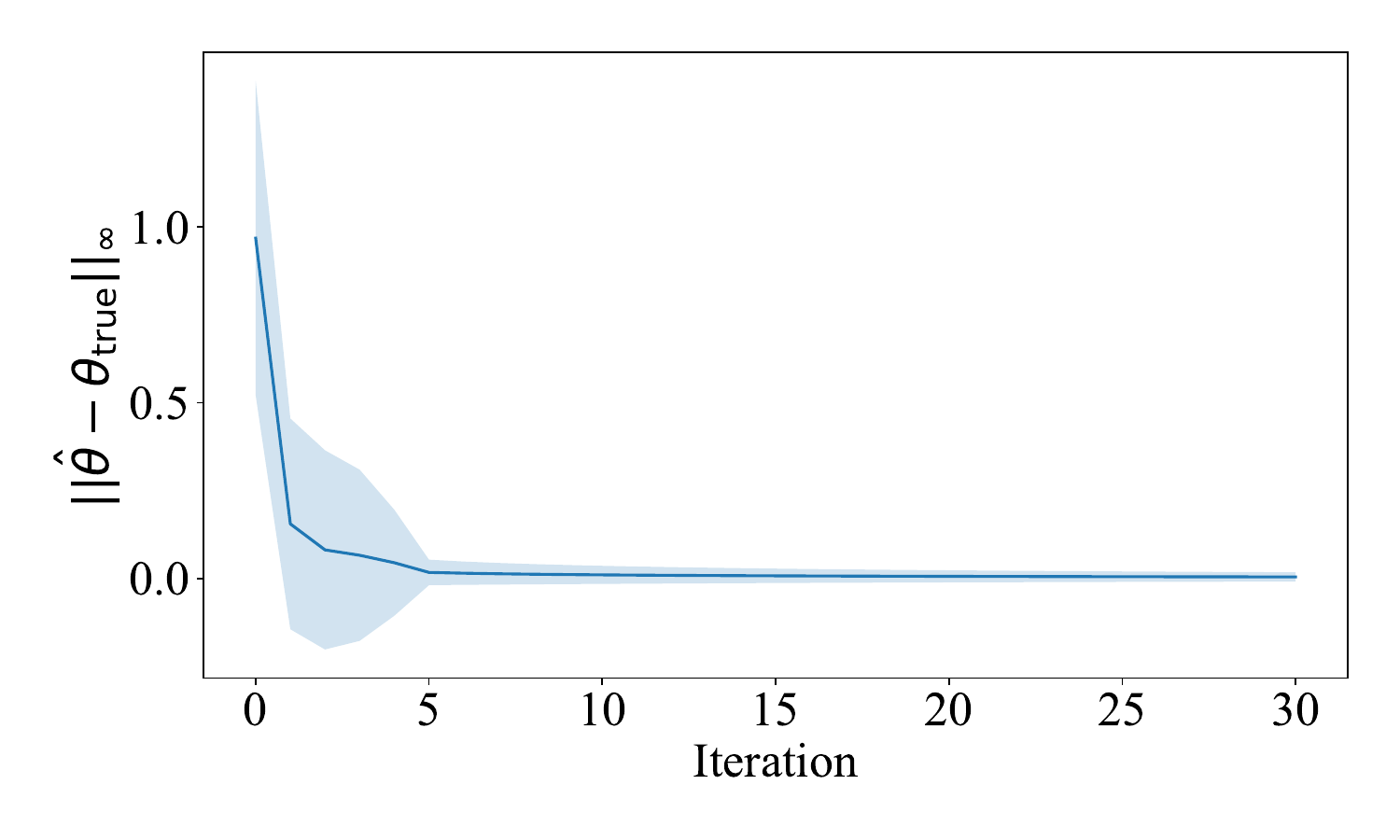}
        \caption{Mean model parameter estimate error and 1-standard deviation interval. The randomness is over $\theta_\mathrm{prior}$, $\Sigma_\mathrm{prior}$ and initial $\Sigma$.}
        \label{fig:unicycle_error}
    \end{subfigure}

    \caption{Results for the unicycle dynamics.}
    \label{fig:unicycle}
\end{figure*}

We also evaluate our algorithm on the unicycle dynamics model, which is nonlinear and given by the equation
\begin{equation}\label{eq:unicycle}
    x_{t+1} = \begin{bmatrix}
        x_{t, 1} + u_{1,t} \Delta t \cos(x_{t, 3})\\
        x_{t, 2} + u_{1,t} \Delta t \sin(x_{t, 3}) \\
        x_{t, 3} + u_{2,t}\Delta t
    \end{bmatrix},
\end{equation}
where $u_{1,t}$ is the linear velocity input and $u_{2,t}$ is the angular velocity input. We use $\Delta t=0.1$. In this example, we will only be able to choose the control input $u_t$ at every timestep. In other words, the informative input design must be done in sequence, with the next state depending on the previous state and input. The parametric model is assumed to be
\begin{equation}
    f(x_t, u_t; \theta)= \begin{bmatrix}
        \theta_2 x_{t, 1} + u_{1,t} \theta_1 \cos(x_{t, 3})\\
        \theta_2 x_{t, 2} + u_{1,t} \theta_1 \sin(x_{t, 3}) \\
        x_{t, 3} + u_{2,t}\theta_1
    \end{bmatrix}.
\end{equation}
The time discretization is thus one of the unknown parameters. \Cref{fig:unicycle} shows the performance of our algorithm. 
Once, $\theta_1$ is correctly estimated, the ellipses become vertical, since there is uncertainty only in $\theta_2$. In this experiment, we bound the norm of the input $u_{1,t}$ to be less than $1$.


\subsection{Model Mismatch}
\begin{figure}[h] 
    \centering
    \includegraphics[width=0.4\textwidth]{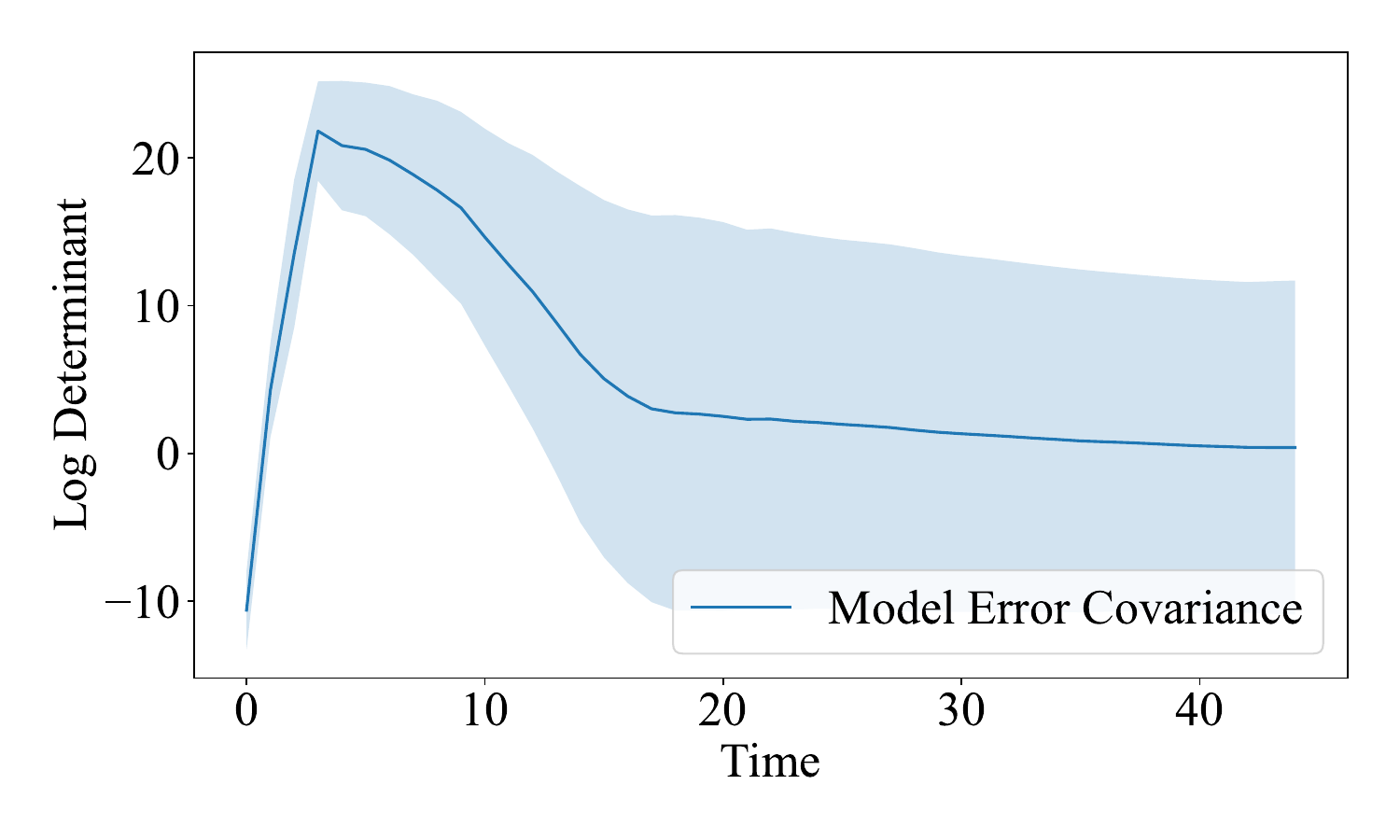}
    \caption{Mean $\log \det \Sigma_\mathrm{model\ error}$ and 1-standard deviation interval over time for the parametric model family \eqref{eq:bad_1} and the Hénon dynamics. The fact that $\log \det \Sigma_\mathrm{model\ error}$ plateaus higher than its initial value indicates that this is an inadequate parametric model family. }
    \label{fig:modelmismatch}
\end{figure}

We consider the case where the parametric model family does not include the true system. We say that the model family does not capture the true dynamics. We suppose that the true system is the Hénon dynamics of \cref{eq:henon} and the parametric model family consists of linear models of the form
\begin{equation}\label{eq:bad_1}
    f(x_t; \theta) = \begin{bmatrix}
        \theta_1 & \theta_2\\
        \theta_1 & \theta_2
    \end{bmatrix}x_t.
\end{equation}
Obviously, this is an inadequate model family. As discussed in \cref{sec:algo}, we expect our algorithm to provide an indication of the inadequacy of the parametric model family via the magnitude of $\Sigma_\mathrm{model\ error}$. This is shown in \Cref{fig:modelmismatch}. The magnitude of $\Sigma_\mathrm{model\ error}$, here chosen as $\log \det \Sigma_\mathrm{model\ error}$ plateaus higher than its initial value. This implies that the realized errors remain high. Lines 4-5 of \Cref{alg:proposed} ensure that the large model errors are not because of a bad step in the optimization of \eqref{eq:taylor_objective}. In other words, lines 4-5 make it unlikely that we have the appropriate model  family, but a bad estimate $\hat{\theta}$ due to the linearization error in problem \eqref{eq:taylor_objective}. This indicates that the obtained parameter estimates, which are in a sense optimal, are inadequate to capture the true dynamics. In the case of the model family \eqref{eq:henon_model}, which captures the dynamics, $\log \det \Sigma_\mathrm{model\ error}$ plateaus lower than its initial value. In other words, we suggest the following empirical guideline:

\textit{If $\log \det \Sigma_\mathrm{model\ error}$ plateaus at a value higher than its initial value and sufficiently large (close to $0$) or rises across time, consider modifying the model family. Equivalently, we can look at the model errors $\epsilon_i^\mathrm{model}$ to draw our conclusions.}

\begin{figure}[h] 
    \centering
    \includegraphics[width=0.4\textwidth]{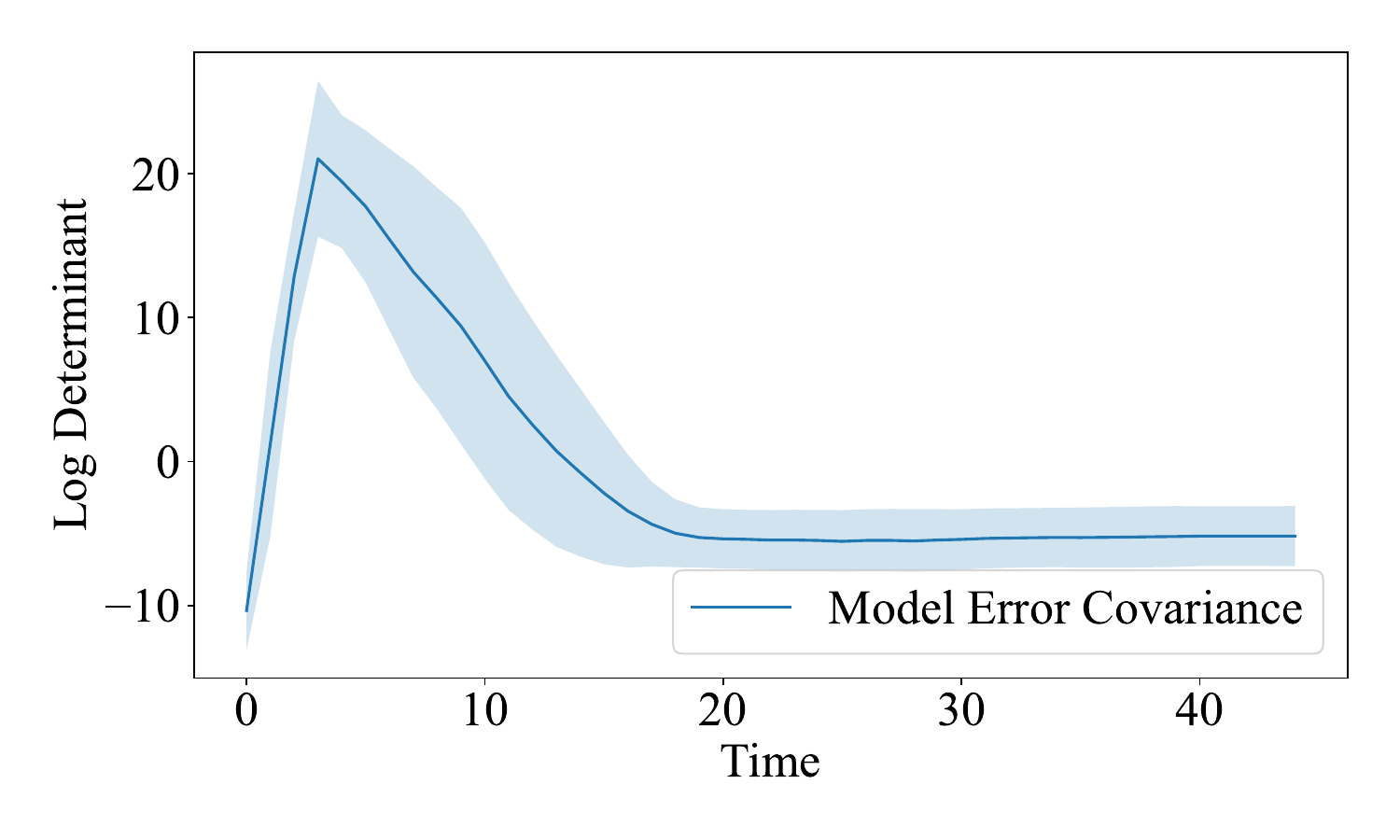}
    \caption{Mean $\log \det \Sigma_\mathrm{model\ error}$ and 1-standard deviation interval over time for the parametric model family \eqref{eq:lin_model} and the Hénon dynamics. The fact that $\log \det \Sigma_\mathrm{model\ error}$ plateaus higher than its initial value indicates that this is an inadequate parametric model family. }
    \label{fig:modelmismatch_2}
\end{figure}

We repeat the process for the inadequate model family \eqref{eq:lin_model} in \Cref{fig:modelmismatch_2}. 

\section{Conclusion}\label{sec:conclusion}
We present a Bayesian methodology for system identification that is based on linear Gaussian assumptions. It extends to general nonlinear system identification \revmykel{through} the Taylor expansion. A key module in our algorithm is the online calibration of the noise covariance that also allows us to judge the suitability of a parametric model family for the underlying system. Our algorithm however has limitations. The use of the trust region, although necessary to ensure the accuracy of the Taylor series, might forbid large changes in the parameters when these are implied by the data. The calibration of the covariance is dependent on the current estimate value, which can slow down performance if the current estimate is not good. Future work will apply our method to high-dimensional problems and look into theoretical guarantees. 
\section*{Acknowledgements}
Toyota Research Institute (TRI) provided funds to assist the authors with their research, but this article solely reflects the opinions and conclusions of its authors and not TRI or any other Toyota entity. The NASA University Leadership Initiative (grant $\#$80NSSC20M0163) provided funds to assist the first author with their research, but this article solely reflects the opinions and conclusions of its authors and not any NASA entity. For the first author, this work was also partially funded through the Alexander S. Onassis Foundation Scholarship program.
We would like to thank Joshua Ott and Arec Jamgochian for their feedback.

\bibliographystyle{IEEEtran}
\bibliography{sample}  

\end{document}